\documentclass[a4paper,12pt]{article} 

\usepackage{gensymb}
\usepackage{physics}
\usepackage{graphicx}
\usepackage{cite}
\usepackage{amsmath}
\usepackage{amssymb}
\usepackage{tabularx}
\usepackage[english]{babel}
\usepackage[left=2cm,right=2cm,top=2cm,bottom=2.5cm,head=2cm,foot=2cm]{geometry}

\newcommand{\der}{\mathrm{d}}

\author{Mátyás Molnár$^{1}$ for the STAR Collaboration,\\
        {\small $^1$Eötvös Loránd University, Budapest, Hungary}}          
\title{Charged particle pseudorapidity distributions measured with the STAR EPD}
\date{\today}

\begin{document}

\maketitle

\begin{abstract}
In 2018, in preparation for the Beam Energy Scan II, the STAR detector was upgraded with the Event Plane Detector (EPD). The instrument enhanced STAR's capabilities in centrality determination for fluctuation measurements, event plane resolution for flow measurements and in triggering overall. Due to its fine radial granularity, it can also be utilized to measure pseudorapidity distributions of the produced charged primary particles, in EPD's pseudorapidity coverage of $2.15<|\eta|<5.09$. As such a measurement cannot be done directly, the response of the detector to the primary particles has to be understood well. The detector response matrix was determined via Monte Carlo simulations and corrected charged particle 
pseudorapidity distributions were obtained in Au+Au collisions at center of mass collision energies $\sqrt{s_{_{NN}}}$ = 19.6 and 27.0 GeV using an iterative unfolding procedure. Several systematic checks of the method were also done.
\end{abstract}

\section{Introduction}
According to quantum chromodynamics, quarks cannot be observed in their free form, only in hadrons due to the color confinement. This effect also causes the strong interaction to have a finite range of around $10^{-15}$ m -- even though the gluon mass is known to be zero. In the very early Universe with enormous pressure and temperature, it is assumed that these particles could exist in a form of quark--gluon plasma (QGP). To create such a state experimentally, particle accelerators that perform high-energy heavy-ion collisions are utilized. Since the lifetime of the QGP is very short, the information about the partonic state has to be deduced from the final-state particles, e.g. hadronic jets.

One of the experimental facilities studying the formation and the evolution of the QGP is the Relativistic Heavy Ion Collider (RHIC) at the Brookhaven National Laboratory, and one of its experiments is the Solenoidal Tracker at RHIC (STAR)~\cite{star}. The complex STAR detector system consists of several instruments; one of them is the Event Plane Detector (EPD)~\cite{starepd}.

In these proceedings, measurements of charged particle~\footnote{The EPD is more sensitive to charged particles, as detailed subsequently.} pseudorapidity distributions in Au+Au collision data at $\sqrt{s_{NN}}=19.6$ and 27.0 GeV utilizing the EPD are presented. Detailed systematic uncertainty checks are also discussed.

\subsection{The EPD}
\label{sec:epd}
The EPD was installed in 2018, as a part of the preparation for the BES-II program. Among motivations behind building the detector were: improving the event plane resolution for flow measurements, independent centrality determination for fluctuation measurements, and using it as a trigger in high luminosity environment during the BES-II program.

The EPD is a completely new subdetector that was supposed to improve the event plane resolution: for example, by about a factor of 2 in Au+Au collisions at $\sqrt{s_{\text{NN}}}=19.6$ GeV~\cite{tlusty2018rhic}. Its predecessor (in event plane determination), the Beam-Beam Counter (BBC) has much less fine granularity than the EPD: only 36 tiles, with the 18 inner smaller tiles used \-- compared to the 372 tiles of the EPD~\cite{starepd}. It also has smaller acceptance of $3.3<|\eta|<5.0$ in pseudorapidity~\cite{bbc_10.1063/1.2888113}.

The detector consists of two ``wheels'' on either (West and East) side of the STAR detector system, installed $\pm 375$ cm from the nominal interaction point (the detector's center). Each wheel consists of 12 ``supersectors'' covering $\phi=30\degree$ in azimuthal angle, each further segmented to 31 ``tiles'', thus giving 16 radial segments so-called ``rings''\footnote{The rings are numbered from 1 to 32 in the following manner: the innermost East EPD ring is the \#1 which follows outerwards until \#16; then, the \#17 continues on the West EPD side's outermost ring until \#32 being the innermost one.} covering relatively large forward pseudorapidity range of $2.15<|\eta|<5.09$ (or, range of $0.7\degree < \theta < 13.5\degree$ angle to particle beam axis). Each supersector is connected to a bundle of 31 optical cables that transport light to high-efficiency silicone photo-multipliers (SiPM). The signals are then sent to the digital data acquisition systems~\cite{starepd}.

\begin{figure}
\centering
\includegraphics[width=0.6\linewidth]{"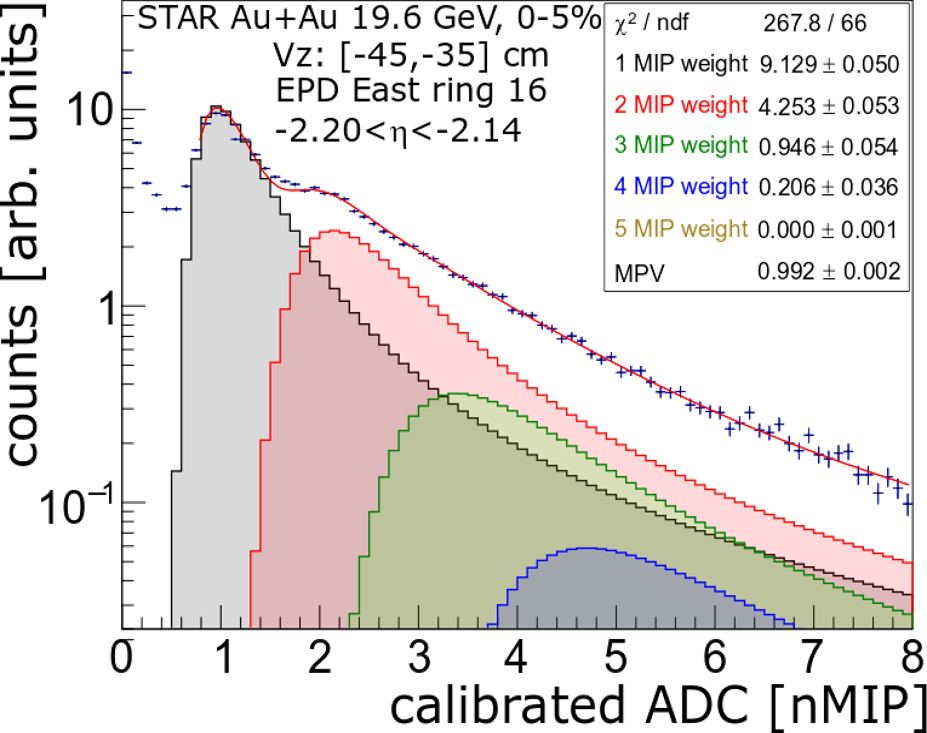"}
\caption{Example multi-MIP Landau fit of ADC count distribution in ring \#16, with ADC counts in arbitrary units. Blue points with error bars represent the data, red continuous line shows the fitted function.}
\label{fig:ADC_mip}
\end{figure}

Each tile registers hits, mostly Minimum Ionizing Particles (MIPs). Assuming that the probability distribution of the measured signal of a single hit can be described by a Landau distribution, the presence of multiple hits will result in a convolution of multiple Landau distributions. The measured Analog Digital Count (ADC) distributions were fitted with a multi-MIP Landau function, shown in Fig.~\ref{fig:ADC_mip}. The different Landau distributions corresponding to the ADC contribution caused by $n$ number of MIPs were convolved with different convolution weights ($n$-MIP weight).

The conclusion drawn was that convolving with less than 5 $n$-MIP weights are adequate to achieve a good fit, as the contribution of the 5-MIP weight was already zero within uncertainties -- under the asssumption that the MIP weights were Poisson-distributed which was validated during data analysis. In view of this result, the systematic uncertainty contribution from this source -- that is, fitting only up to 5 $n$-MIP weights -- can be considered negligible.

\unskip

\section{Methodology}
\subsection{Charged particle pseudorapidity measurement with the EPD}
The aim is to measure the angle $\theta$ between the three-momentum $\vb{p}$ of the particle and the beam. Instead, a more convenient\footnote{In the ultrarelativistic limit, it approaches to rapidity (in $c=1$ unit system, $c$ being the speed of light): $\eta\approx y\equiv\frac12\ln\left(\frac{E+p_z}{E-p_z}\right)$, with $E$ being the energy of the particle.} quantity, the pseudorapidity $\eta$ is used, which is defined as:
\begin{equation}
	\eta \equiv -\ln\left[\tan\left(\frac{\theta}{2}\right)\right]=\frac{1}{2}\ln\left(\frac{|\vb{p}|+p_z}{|\vb{p}|-p_z}\right),
\end{equation}
where $p_z$ is the $z$ component of the momentum, and the $z$ direction is chosen to coincide with the direction of the beam~\cite{pseudorapidity}.

Beyond the event plane determination, the EPD's fine radial granularity allows for pseudorapidity measurements to be performed. The raw EPD hit numbers could be used to calculate the pseudorapidity distribution of charged particles ($\der N_{ch}/\der\eta$) by using the corresponding $\eta$ value of the given ring.

However, this also includes the secondary particles that do not originate from the primary vertex. As the EPD is preceded by the rest of the detector system and is relatively far from the interaction point, multiple factors distort (``blur'')  the measured distribution.

\begin{figure}
	\begin{minipage}[t]{0.5\textwidth}
	\centering
	\includegraphics[width=0.95\linewidth]{"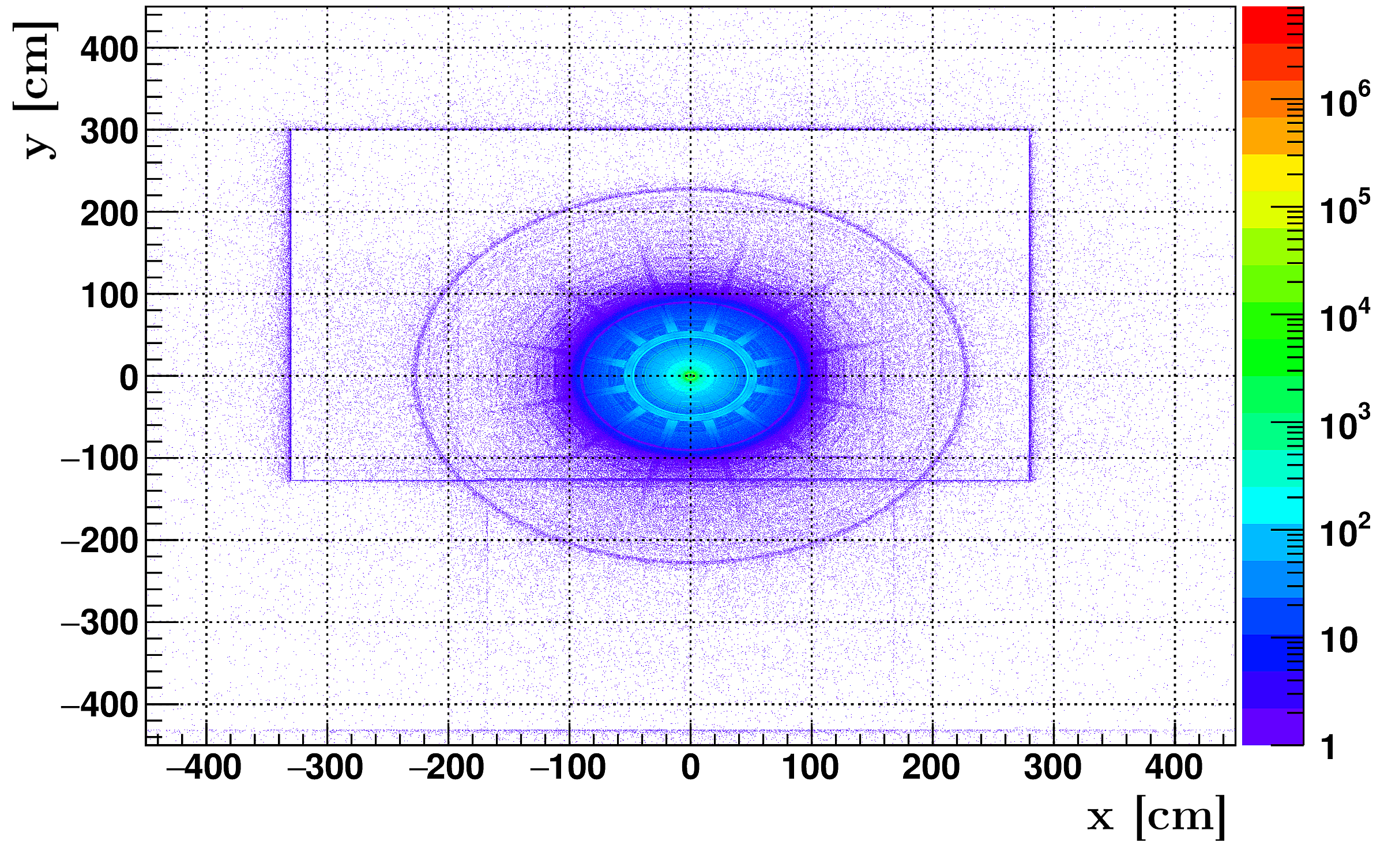"}\\
	a)
	\end{minipage}
	\hfill
	\begin{minipage}[t]{0.5\textwidth}
	\centering
	\includegraphics[width=0.95\linewidth]{"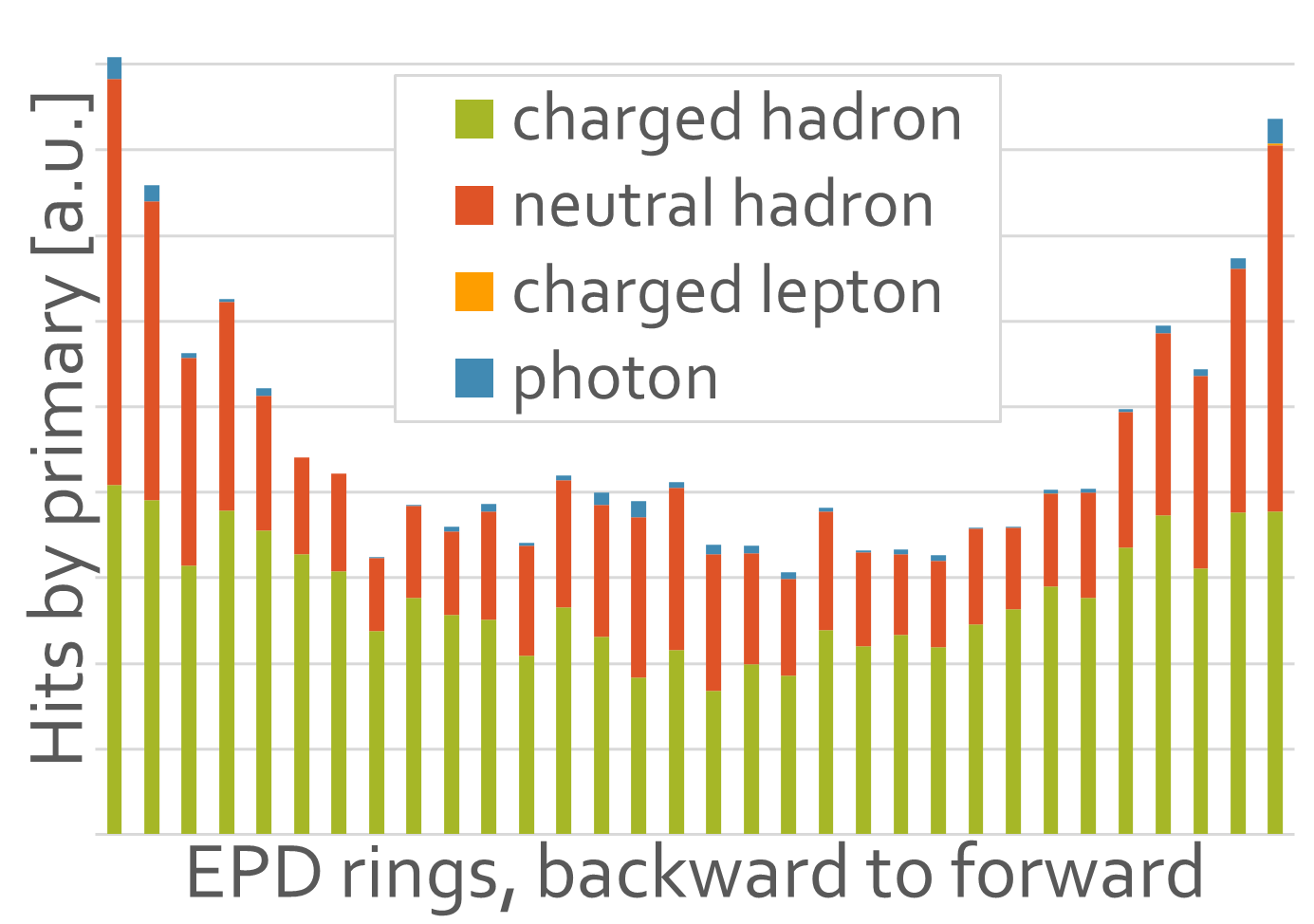"}\\
	b)
	\end{minipage}
\caption{(\textbf{a}) Vertices of particles registered by the EPD, based on a \texttt{HIJING}~\cite{hijing} + \texttt{Geant4}~\cite{geant} MC detector simulation. The plots shows the vertex distribution in the $x$--$y$ plane, integrated along the $z$ axis, revealing the detector structure and surrounding materials.  (\textbf{b}) Distribution of various types of simulated primary particles hitting EPD, ring-by-ring, where rings in the backward direction are in the left hand part of this panel, while rings in the forward direction are in the right hand side -- ordered by apparent spatial rapidity of the given ring.}
\label{fig:secondary}
\end{figure}

The factors assumed to cause the most significant distortion effect are as follows. First of all, charged primary particles scatter in detector material (or in rare cases with each other), creating secondary particles contributing to $\der N_{\textnormal{ch}}/\der\eta$ significantly. This is demonstrated in Fig.~\ref{fig:secondary}a, where the vertices (origins) of particles hitting the EPD in a detector material simulation are depicted. Secondly, neutral primary particles contribute through decays (e.g. a neutral $\Lambda$ baryon decaying into proton and pion). In Fig.~\ref{fig:secondary}b it is clearly demonstrated that this contribution is non-negligible (based on the same simulation as mentioned above).

\subsection{From raw EPD data to pseudorapidity distribution [$\der N/\der\eta$]}
Using the previously mentioned multi-MIP Landau fit, one can extract the number of EPD hits for each ring; denoted as $N(i_{\text{Ring}})$ in the $i$-th ring. Given the underlying pseudorapidity distribution of the primary particles ($\der N/\der\eta$), assuming linear dependence from the $\der N/\der\eta$, the number of hits in a given ring can be calculated formally as a convolution:
\begin{equation}
	\der N(i_{\text{Ring}}) = \int R(\eta, i_{\text{Ring}})\frac{\der N}{\der\eta}\der\eta,
\end{equation}
where $R$ denotes the \textit{response matrix}, which encodes response of the detector, i.e. connects a detector-level distribution with the true distribution to be measured. In this analysis, it contains the number of hits in the given ring number distribution's bin, originating from a particle at given $\eta$ pseudorapidity distribution's bin.

No probabilistic consideration guarantees this matrix to be invertible, therefore a simple (or even a regularized) matrix inversion might not be an option even if the exact form of $R$ would be known. Instead, a method called \textit{bayesian iterative unfolding}~\cite{unfold_theor} (``deblurring'') is used.

Using this approach, the $R$ needs to be extracted from simulations that are as close to the real system as possible. Using a complex event generator, a list of primary particles is obtained, along with a list of EPD hits -- preferably all linked to primary tracks causing them.

\begin{figure}
\centering
\includegraphics[width=0.6\linewidth]{"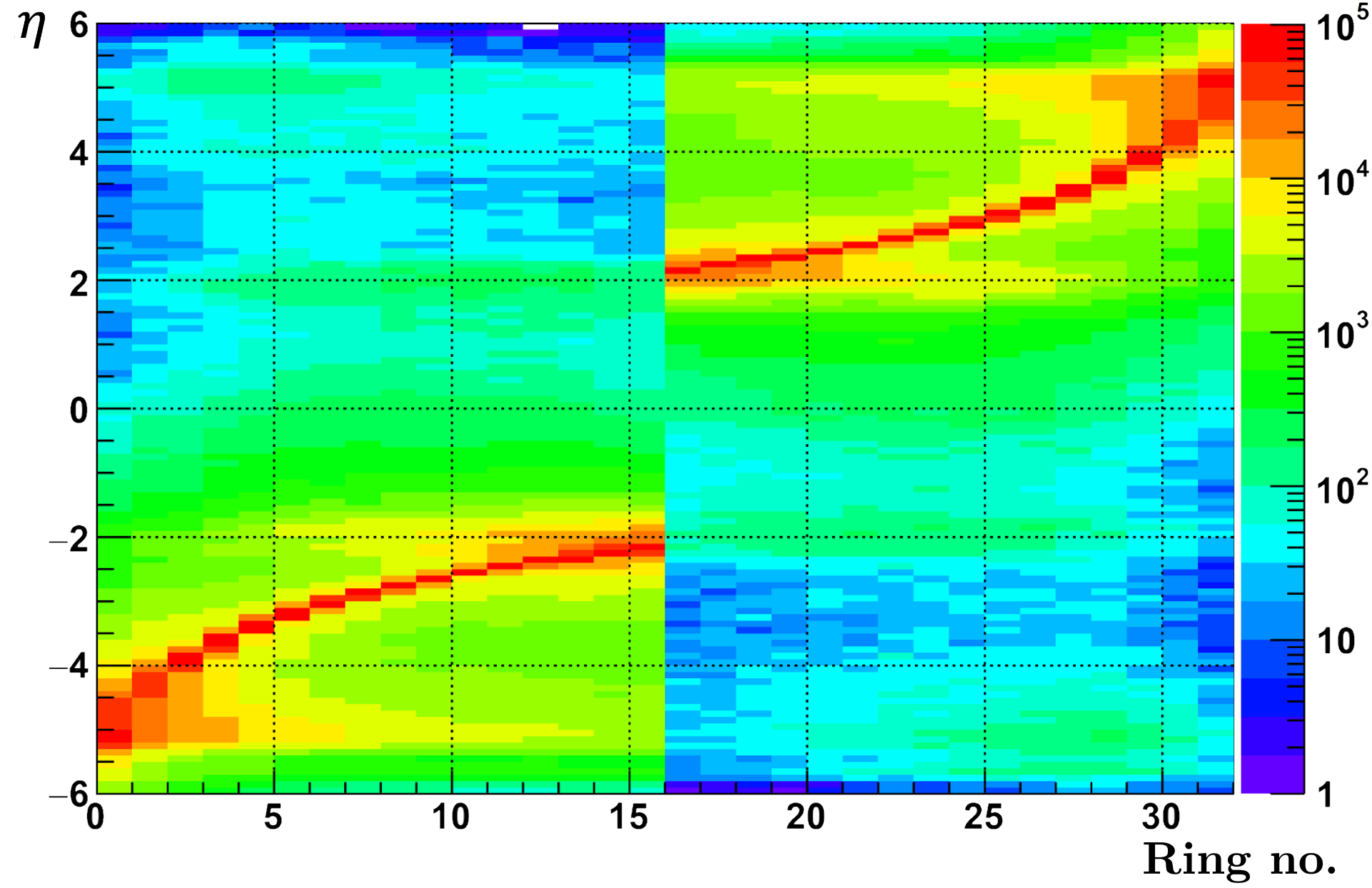"}
\caption{Heatmap visualization of the $R$ response matrix, connecting bins containing numbers of EPD ring hits (caused by either primary or secondary particles) with bins corresponding to primary particles at given $\eta$ pseudorapidity. The left side corresponds to East EPD wheel, the right side to West EPD wheel. It is worth noting that many primaries create hits even in the opposite side EPD via secondaries, as seen in upper left and bottom right quarters.}
\label{fig:response_mx}
\end{figure}

In this analysis, the events were generated using the STAR's \texttt{HIJING} Monte Carlo event generator combined with \texttt{Geant4} to simulate the precise geometry of the EPD. In the following, the abbreviation MC will indicate data from these simulations. Such a response matrix can be seen in Fig.~\ref{fig:response_mx}.

It should be noted that no (light) ion fragments can be simulated in \texttt{HIJING}, which are, in reality, inevitable with heavy-ion collisions. However, this shortfall should not change the results significantly, according to PHOBOS results~\cite{phobos_fragments}: the contribution from light ion fragments causes at least an order of magnitude smaller contribution to $\der N/\der\eta$ than the results in this analysis (see Sec. \ref{sec:results}).

In the following step, the unfolding technique is utilized to determine an uncorrected d\textit{N}/d$\eta$. The software used for this purpose is the \texttt{RooUnfold}~\cite{roounfold} framework, implemented in C++, running within the \texttt{ROOT} environment~\cite{root}.  The package itself defines classes for the different unfolding algorithms -- amongst others, the bayesian iterative unfolding.

The response matrix class of the software includes functions for populating the response matrix\footnote{\texttt{Fill($x_{\text{measured}}$, $x_{\text{truth}}$)}; naturally, ``measured'' and ``truth'' here stand for the training datasets obtained from MC (simulation).} as well as for managing the background (missed hits from real primaries and hits resulting from other sources\footnote{\texttt{Miss($x_{\text{truth}}$)} and \texttt{Fake($x_{\text{measured}}$)}}).

During the unfolding, one can choose to propagate the statistical uncertainty in different ways; in this case, the most appropriate method should be propagating the (mostly badly conditioned, thus non-invertible) covariance matrix~\cite{unfold_theor}.

The resulting EPD ring distribution~\footnote{Caused by both primary and secondary particles.} needs to be corrected for the \textit{multiple counting} (efficiency, $\epsilon$), explained as follows. The unfolding procedure results in one unfolded track for each individual EPD hit. However, it should be noted that one primary track can cause multiple hits. This effect needs to be corrected for -- either via a bin-by-bin correction calculated from MC data (via a \texttt{Number of hits from 1 primary}($\eta$) distribution), or by weighing the values filled in response matrix such that it could compensate for the multiple counts during the unfolding. In this analysis, the first method was used.

\subsection{Extracting charged particle pseudorapidity distribution}
\label{sec:chargedfactor}
In order to obtain the charged particle distribution ($\der N_{ch}/\der\eta$) from $\der N/\der\eta$, either different bin-by-bin corrections can be used, or neutral particles can be marked as background (``fake'') using RooUnfold's \texttt{Fake()} method. In this analysis, the following methods were used as the \textit{charged factor correction}:
\begin{enumerate}
	\item Bin-by-bin correction of the already unfolded $\der N/\der\eta$ using the charged particle fraction $N_{charged}(\eta)/N_{all}(\eta)$ from MC data;
	\item Bin-by-bin correction of the raw EPD data via $N_{charged}(i_{\text{Ring}})/N_{all}(i_{\text{Ring}})$ from MC data; then unfolding of the EPD charged particle distribution.\footnote{In this case, another type of response matrix has to be used that was filled only with the charged particles' data.}
	\item Mark neutral particles as background and fill the response matrix as in the second method, except that the hits from neutral primaries are considered as ``fake''.
\end{enumerate}

The three different methods can later be used to estimate the systematic uncertainty of the unfolding procedure itself.

\subsection{Consistency check of the unfolding methods}
\label{sec:closure}
Before unfolding the real data, a closure test was done to check whether the unfolding method can recover the ``true'' training data itself (MC ``truth''). 

It was found that unfolding done on the input training MC sample reproduces well the input $\eta$ distribution. When some noise ($\pm1$--$10\%$) was added to the training sample, the resulting unfolded distribution was in agreement with the input distribution within $<3\%$. All in all, the unfolding itself was found to work well.

\begin{figure}
\centering
\includegraphics[width=0.6\linewidth]{"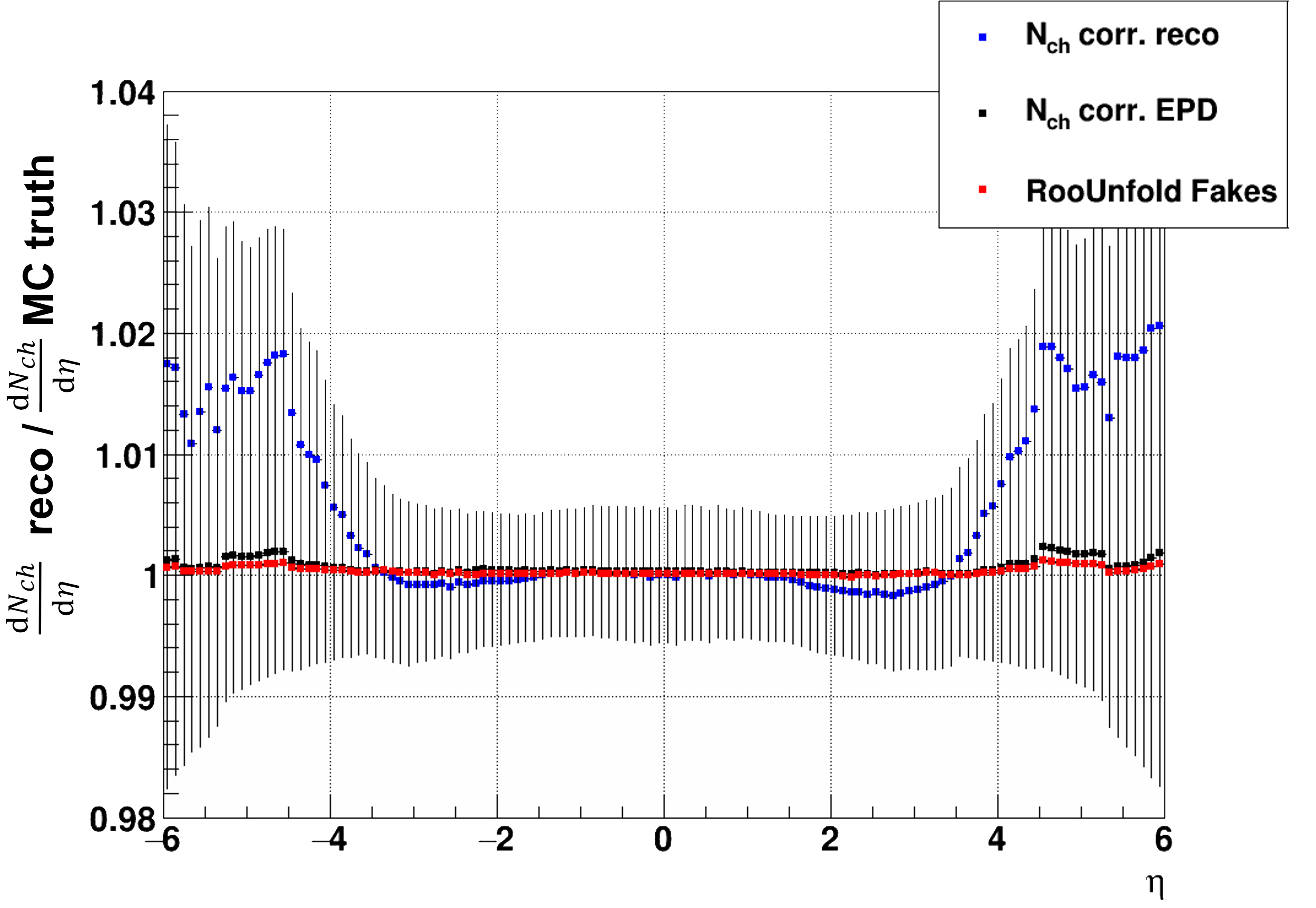"}
\caption{Consistency check of the three different methods to get $\der N_{ch}/\der\eta$ from MC EPD ring distribution. The difference is shown as unfolded $\der N_{ch}/\der\eta$ over MC ``truth'', the distributions divided bin-by-bin. Blue marker represents the first method ($\eta$-dependent charged factor correction), black shows the second method (EPD ring number dependent charged factor correction), and red represents the third method (marking neutral particles), relative to MC truth's $\der N_{ch}/\der\eta$. The errorbars are only plotted for informative purposes: they were calculated using the \texttt{ROOT}'s \texttt{TH1} class' default square root of sum of squares of weights.}
\label{fig:relat_closure_test}
\end{figure}

Furthermore, after applying the multiple counting correction and the three different methods of charged factor correction on the unfolded distribution\footnote{Note that the mentioned unfolding procedure was at this stage still done on the MC EPD ring distribution, thus, on the training sample.}, the resulting distributions were compared to the original MC dataset's $\der N_{ch}/\der\eta$. As it is visible in Fig.~\ref{fig:relat_closure_test}, the maximal relative deviation is up to 2\% in certain bins for the first method and less than 0.1\% for the other two methods.

It is worth noting that although the third method (marking neutral particles) shows here the most precise result, the systematic checks showed that it is the least reliable, in terms of most heavily depending on the MC input provided to the response matrix.

Given the result of the closure test, the unfolding and correction methods were considered adequately self-consistent.

\section{Systematic checks}
\label{sec:systematics}
In the following section, the examined systematic uncertainty sources and their contribution to the results are discussed.

\subsection{Dependence on input MC distribution}
The bayesian iterative unfolding process, via its iterative nature, should mostly overcome differences in response matrix from real response that are not related to distortion effects, such as detector geometry~\cite{unfold_theor}. However, as the exact response matrix cannot be determined even with precise MC simulations and the unfolding process itself is not perfect, some dependencies on the various parameters in the MC simulations can occur. Those are considered as systematic uncertainties of the measurement.
\subsubsection{Tightening and shifting the input MC $\der N/\der\eta$}
\label{sec:narrow}
\begin{figure}
	\centering
	\hspace*{1.7cm}
	\includegraphics[width=0.75\linewidth]{"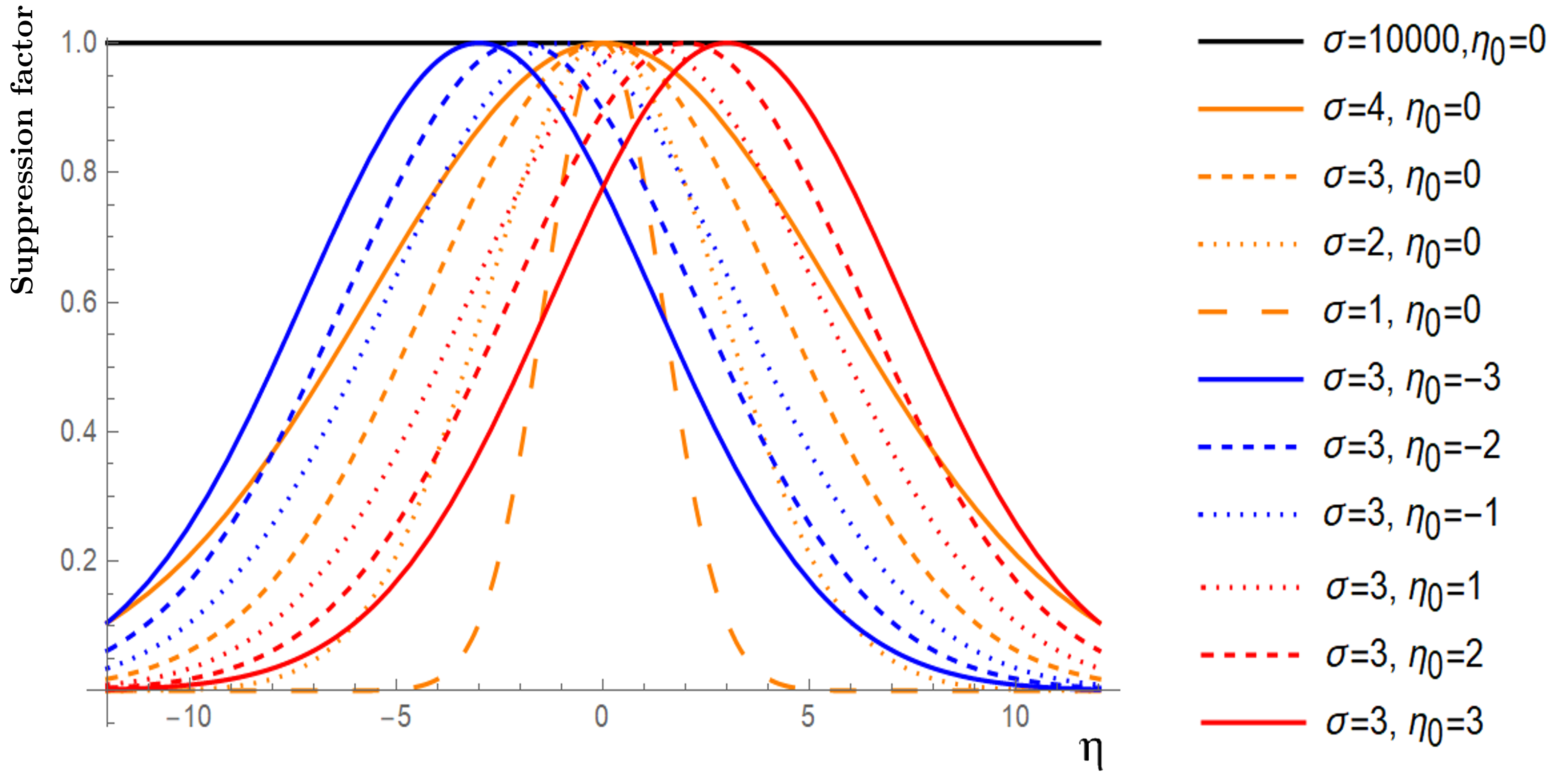"}\\
	a)\\
	\includegraphics[width=0.65\linewidth]{"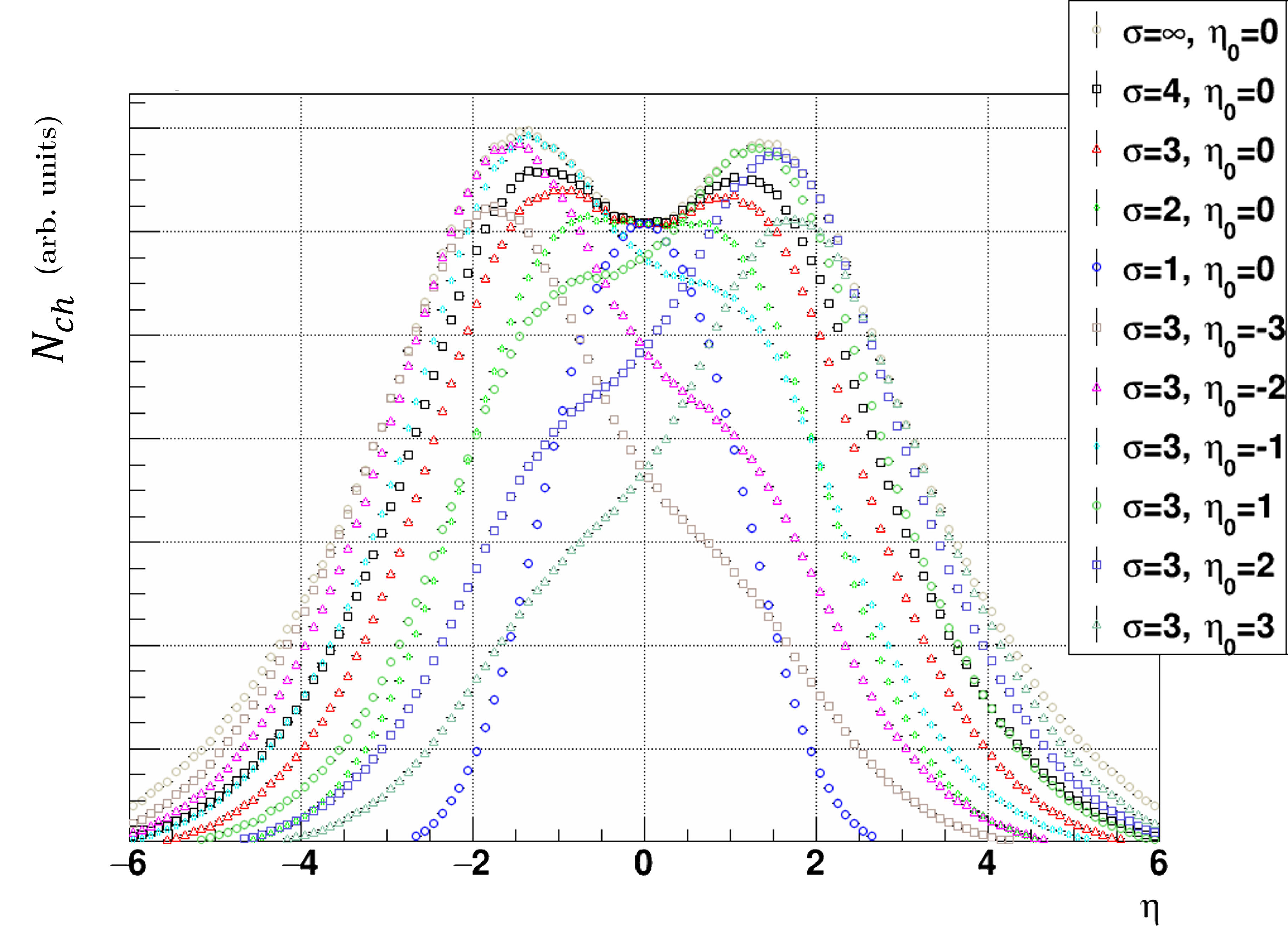"}\\
	b)
\caption{Tightening and shifting the MC input distribution using random selection based on Gaussian distribution of $\sigma$ width and $\eta_0$ curve peak position. (\textbf{a}) Demonstration of the Gaussian suppression factors used. (\textbf{b}) The $\der N_{ch}/\der\eta$ of the distorted MC input samples.}
\label{fig:mctight}
\end{figure}

Firstly, the simulated sample's $\der N_{ch}/\der\eta$ was modified (``suppressed'') using a Gaussian shape with width $\sigma$ and mean $\eta_0$. These suppression factors can be seen in Fig.~\ref{fig:mctight}a. This was done via a random selection based on Gaussian distribution while filling the response matrices.

Using this approach, all combinations could be analysed, that is, unfolding the $i$-th MC sample's EPD ring hit distribution via response from $j$-th MC sample. In case of $i=j$, the unfolding was as close to perfect as expected, discussed in Subsec. \ref{sec:closure}.

Unfolding results with the Gaussian width of $\sigma\lessapprox 1$ were not considered here as in this case there are almost no particles in the EPD range. Otherwise, there was less than a few percent variation in the EPD's $\eta$ region.

Overall, in the analysis the effect of tightening the $\der N_{ch}/\der\eta$ of the training sample to $\sigma=2$ and shifting it by $\pm 3$ units of pseudorapidity was investigated.

\subsubsection{Broadening the input MC $\der N/\der\eta$}
\begin{figure}
	\centering
	\includegraphics[width=0.7\linewidth]{"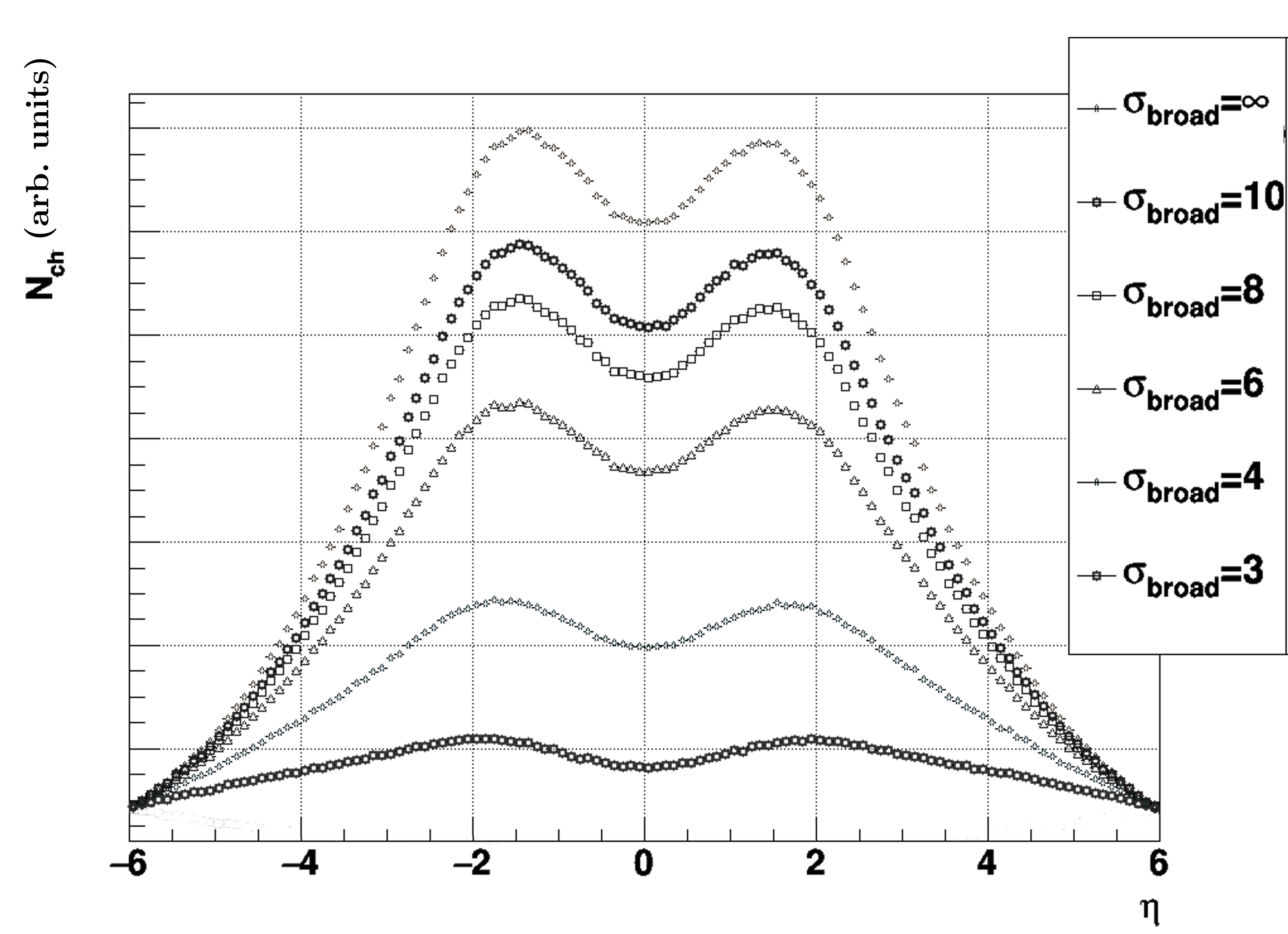"}\\
\caption{Broadening the MC input distribution using random selection based on Gaussian distribution of $\sigma_{\text{broad}}$ width.}
\label{fig:mcbroad}
\end{figure}

Similar to modification done in Sec. \ref{sec:narrow}., here the tracks were modified with a factor of 
\begin{align}
\exp\left(\frac{\eta^2-\eta_\text{max}^2}{2\sigma_{\text{broad}}}\right).
\end{align}
There was no suppression utilized for $|\eta|>\eta_{\text{max}}$, with $\eta_{\text{max}}=6$. The resulting shape of the distributions can be seen in Fig.~\ref{fig:mcbroad}.

While unfolding the data with these input MC distributions, a significant decrease at midrapidity values was observed. However, this occurred mostly outside the EPD's $\eta$ region; the unfolding was considered acceptable down to $\sigma_{\text{broad}}\approx 3$.  

\subsection{Changing the charged fraction in the MC training dataset}
The fraction of the charged particles in the MC input data was changed by $\pm 15\%$. This was achieved by randomly rejecting either the neutral or the charged particles.

\subsection{Changing the $p_{\textnormal{T}}$ slope of the MC training dataset}
The transverse momentum ($p_{\textnormal{T}}$) distribution slope of the MC input data was changed by $\pm 10\%$ via randomly rejecting particles of small or large $p_{\textnormal{T}}$.

\subsection{Centrality and z-vertex selection}
It was investigated, by how much the unfolded distribution would change if either the $z$-vertex or the centrality selection are modified. For the former investigation, a $\pm 5$ cm calibration uncertainty in the z-vertex measurement of the real EPD data was employed; 
for the second one, $\pm 5\%$ calibration uncertainty was assumed in centrality determination of the real EPD data.

\subsection{z-vertex choice}
Due to the detector geometry, it is important to also take into account the interaction point's z-vertex position in the calculations, as the resulting pseudorapidity distribution should not depend on it. 

The EPD data, as well as the responses, were collected in nine different z-vertex classes, equally distributed from $-45$ to $+45$ cm. Depending on which range was unfolded, the resulting distribution still may differ and has to be taken into account as systematic uncertainty.

\subsection{Unfolding method choice}
The most significant systematic uncertainty contribution was caused by the difference between the results achieved using different unfolding and correction methods (as listed in Subsection~\ref{sec:chargedfactor}.). The first method was used as benchmark, from which the differences were calculated.

\subsection{EPD related uncertainties}
As previously stated, the EPD electronics were considered fully efficient (except some ``dead areas'' in the detector from e.g. glue and gaps, but these were assumed to be correctly handled in the simulation). The uncertainty from multi-MIP Landau fit was considered negligible compared to other systematic sources.

In conclusion, the systematic uncertainties coming from the detector system itself were considered negligible.

\begin{table} 
\caption{Summary of systematic uncertainty sources and their contribution.\label{tab:systematics}}
\newcolumntype{C}{>{\centering\arraybackslash}X}
\begin{tabularx}{\textwidth}{lC}
\hline
\textbf{Source}	& \textbf{Systematic uncertainty}\\
\hline
MC input $\der N_{\text{ch}}/\der\eta$ tightening, shifting		& 6\%			\\
MC input $\der N_{\text{ch}}/\der\eta$ broadening		& 4\%			\\
Charged fraction in MC & 6\% \\
$p_{\textnormal{T}}$ slope change in MC & 1\% \\
Centrality selection & 2\% \\
z-vertex selection & negligible \\
z-vertex choice & 1\% \\
Unfolding method choice & 8\% \\
EPD related uncertainties, electronics, efficiency & negligible \\
\hline
\end{tabularx}
\end{table}

The different systematic uncertainty sources and their contribution with informative percentage values can be seen in Table~\ref{tab:systematics}.

\section{Results}
\label{sec:results}

\begin{figure}
	\centering
	\includegraphics[width=0.8\linewidth]{"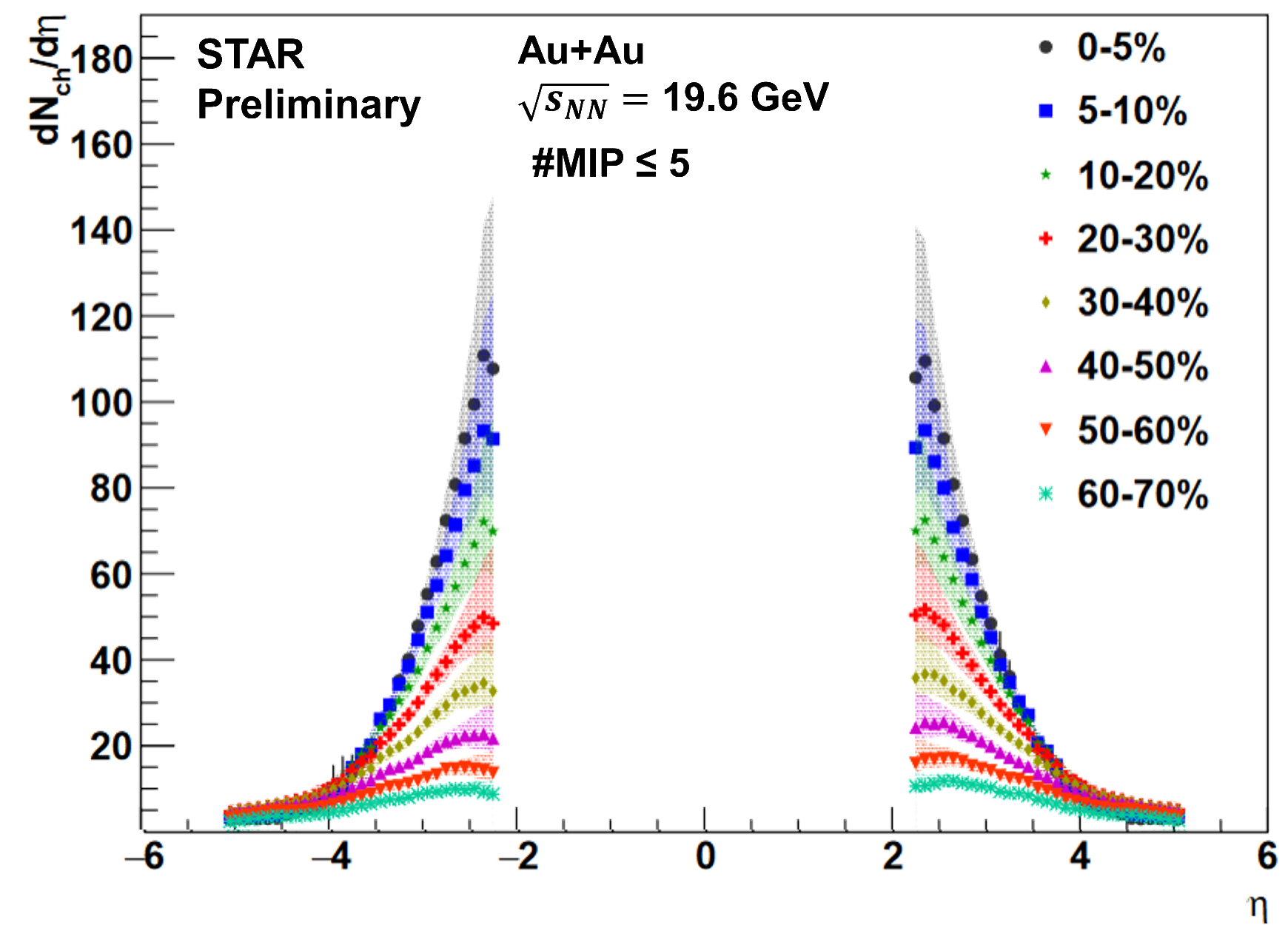"}\\
\caption{Charged particle pseudorapidity distributions measured with STAR EPD on RHIC energy $\sqrt{s_{NN}}=19.6$ GeV. The data was processed in eight centrality classes, presented with the different markers. The statistical uncertainties, marked by errorbars, are not visible on this plot, as the markers themselves are larger. The coloured area indicates the systematic uncertainties of the measurement.}
\label{fig:result20}
\end{figure}

\begin{figure}
	\centering
	\includegraphics[width=0.8\linewidth]{"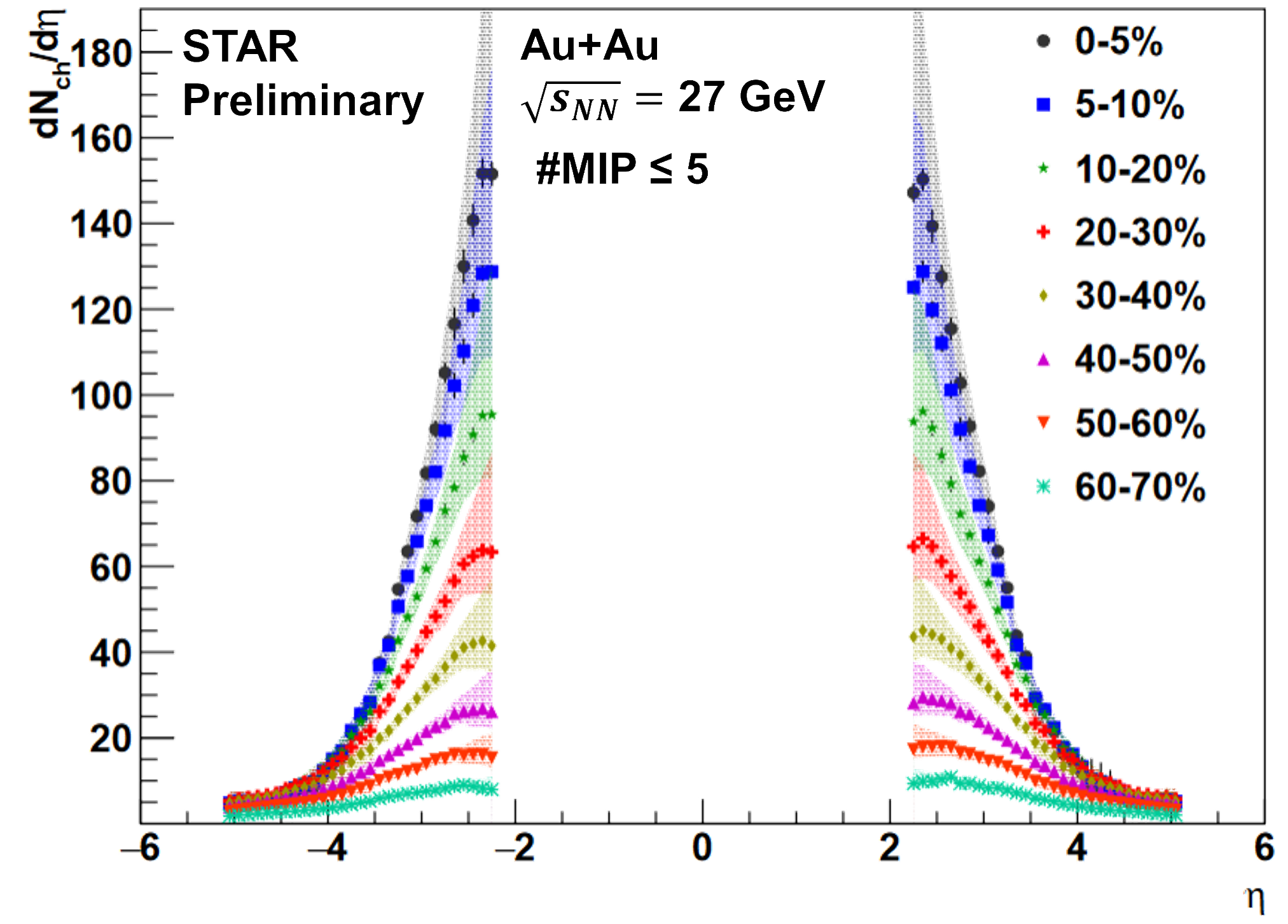"}\\
\caption{Charged particle pseudorapidity distributions measured with STAR EPD on RHIC energy $\sqrt{s_{NN}}=27.0$ GeV. The data were processed in eight centrality classes, presented with the different markers. The errorbars represent the statistical uncertainty, the coloured area indicates the systematic uncertainties of the measurement.}
\label{fig:result27}
\end{figure}

In this manuscript, charged particle pseudorapidity distributions with systematic uncertainties listed in Sec.~\ref{sec:systematics}. were obtained at two RHIC energies, in the EPD pseudorapidity range. The results at $\sqrt{s_{NN}}=19.6$ and 27.0 GeV can be seen in Fig. \ref{fig:result20} and \ref{fig:result27}, respectively. The caption $\#\text{MIP}\leq5$ written on the plot indicates the number of convolution members in the multi-MIP Landau fit, as described in Sec.~\ref{sec:epd}.

\subsection{Comparison with the PHOBOS results}
Another experiment of the RHIC complex was the PHOBOS experiment, which completed data taking in 2006. The PHOBOS was a large acceptance silicon detector, covering almost $2\pi$ in azimuth and $|\eta|<5.4$ in pseudorapidity~\cite{phobos_fragments}. Compared to STAR’s EPD, there are differences in both detector topology and granularity: the silicone pad detectors measure the total number of charged particles emitted in the collision, with modules mounted onto a centrally located octagonal frame (Octagon) covering $|eta|\leq 3.2$, as well as three annular frames (Rings) on either side of the collision vertex, extending the coverage out to $|\eta|\leq 5.4$~\cite{back2005phobos}.

The PHOBOS also measured $\der N_{\text{ch}}/\der\eta$ at 19.6, 62.4, 130, 200 GeV energies~\cite{phobos}. Although in that paper a slightly different centrality binning was used (0--3\%, 3--6\% and 6--10\% instead of 0--5\% and 5--10\%; the other centrality classes were the same), at 19.6 GeV the results can be compared.

\begin{figure}[h!]
	\centering
	\includegraphics[width=0.85\linewidth]{"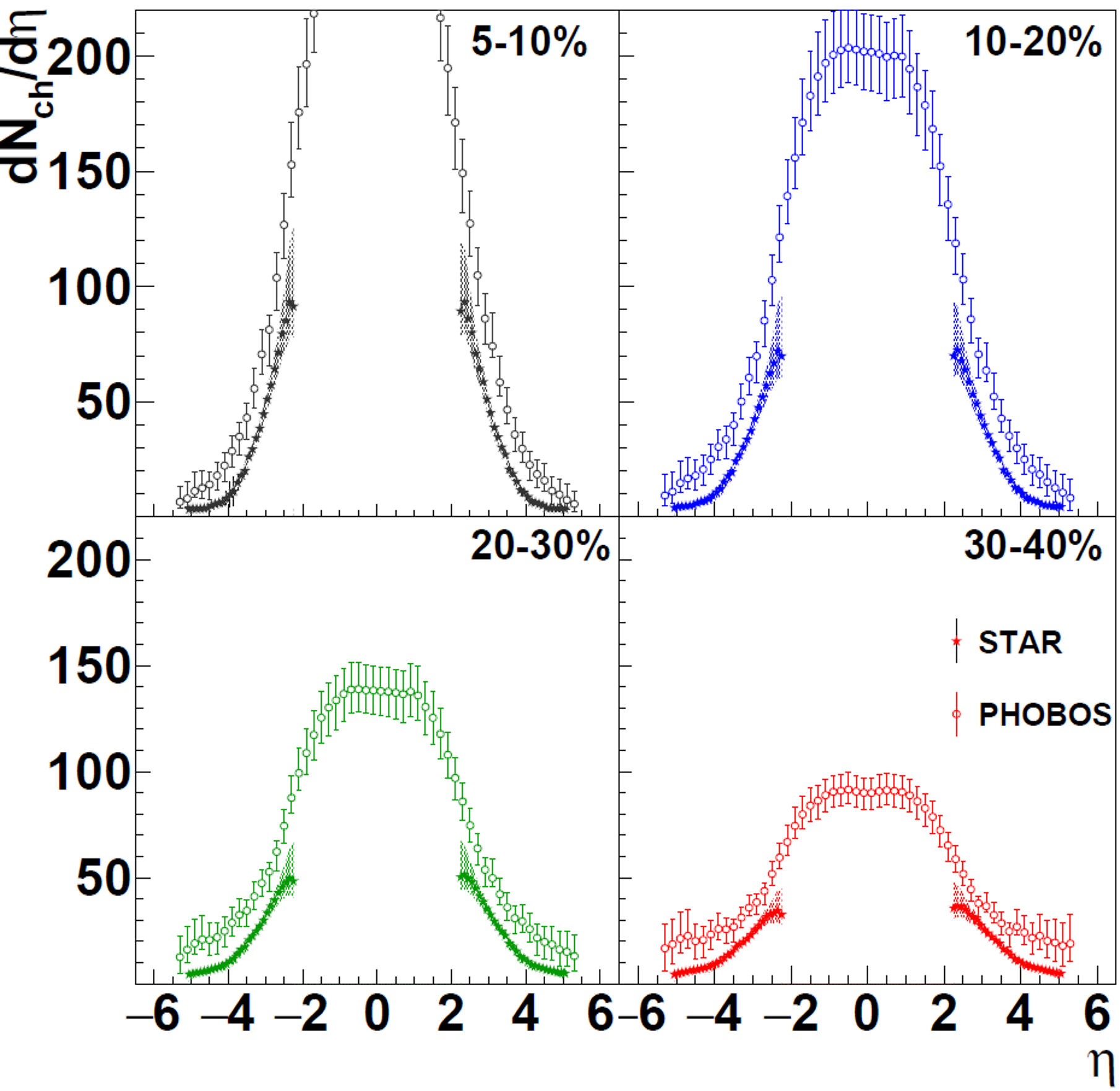"}\\
\caption{Charged particle pseudorapidity distributions measured in PHOBOS (hollow circles) and STAR (star markers). Note that on the upper left graph the centrality class of the PHOBOS experiment's result is actually 6--10\%.}
\label{fig:phobos}
\end{figure}

In Figure~\ref{fig:phobos}, it is apparent that the two measurements show sizeable differences, depending on $\eta$: around up to a factor of two, increasing from small $|\eta|$ towards forward/backward rapidities.

The exact reasons behind this discrepancy are not yet known but the difference cannot be explained by the systematic uncertainties described in Sec.~\ref{sec:systematics}.

\section{Discussion}
In summary, based on EPD ring-by-ring distributions, charged particle pseudorapidity measurements at $\sqrt{s_{\textnormal{NN}}}=19.6$ and 27.0 GeV were performed with detailed systematic investigations regarding simulation data, calibration data, and unfolding methods.

The results at $\sqrt{s_{NN}}=19.6$ GeV show significant difference compared to the results from PHOBOS. There are four components in this comparison: EPD spectrum measurement, \texttt{Geant4} simulation, unfolding procedure from the STAR part, and the PHOBOS data itself.

The method presented in this manuscript is to be extended to other $\sqrt{s_{NN}}$ values (as part of the BES-II program) and to fixed target data -- mainly at energies where the QCD critical point is expected \cite{qcd_crit}. Refining this measurement method is also important for the search of the QCD critical point, in order to fine-tune the models used in these analyses. 

Measuring pseudorapidity values of charged particles is important due to the possibility of estimating the initial energy density of the quark--gluon plasma created in the collisions, based on them~\cite{Csanad:2016add,Ze-Fang:2017ppe}. Furthermore, the forward and backward rapidity measurements can provide information about the nuclear-matter effects as well \cite{forward_nuclear}.

\section*{Acknowledgments}
This research was supported by the NKFIH OTKA K-138136 grant.

\end{document}